\documentstyle[aps,epsfig]{revtex}
\begin{document}
%%%%%%%%%%%%%%%%%%%%%%%%%%%%%%%%%%%%%%%%%%%%%%%%%%%%%%%%%%%%%%%%%%%%%%%%%%%%%%
%%%%%%%% the following newcommands will be completed by the publisher %%%%%%%%
%%%%%%%%%%%%%%%%%%%%%%%%%%%%%%%%%%%%%%%%%%%%%%%%%%%%%%%%%%%%%%%%%%%%%%%%%%%%%%
\newcommand{\volume}{10}              %sets current volume,
\newcommand{\xyear}{2001}            %sets year in header
\newcommand{\issue}{XX}               %sets current issue,
\newcommand{\recdate}{27 June 2001}  %sets received date,
\newcommand{\revdate}{dd.mm.yyyy}    %sets revised date,
\newcommand{\revnum}{0}              %number of revisions,
\newcommand{\accdate}{dd.mm.yyyy}    %sets accepted date,
\newcommand{\coeditor}{ue}           %sets (co)editor,
\newcommand{\firstpage}{1}         %first page number,
\newcommand{\lastpage}{26}          %last page number,
\setcounter{page}{\firstpage}        %sets page counter to first page number
%%%%%%%%%%%%%%%%%%%%%%%%%%%%%%%%%%%%%%%%%%%%%%%%%%%%%%%%%%%%%%%%%%%%%%%%%%%%%%
%%%%%%%%%%%%%%%%%%%%%%%%%%%%%%%%%%%%%%%%%%%%%%%%%%%%%%%%%%%%%%%%%%%%%%%%%%%%%%
%%%%%%%%%%%%%%%%%% please give up to three keywords here %%%%%%%%%%%%%%%%%%%%%
%%%%%%%%%%%%%%%%%%%%%%%%%%%%%%%%%%%%%%%%%%%%%%%%%%%%%%%%%%%%%%%%%%%%%%%%%%%%%%
%\newcommand{\keywords}{atomic force microscope, atomic orbitals}
%%%%%%%%%%%%%%%%%%%%%%%%%%%%%%%%%%%%%%%%%%%%%%%%%%%%%%%%%%%%%%%%%%%%%%%%%%%%%%
%%%%%%%%%%%%%%%% please give up to three PACS numbers here %%%%%%%%%%%%%%%%%%%
%%%%%%%%%%%%%%%%%%%%%%%%%%%%%%%%%%%%%%%%%%%%%%%%%%%%%%%%%%%%%%%%%%%%%%%%%%%%%%
\newcommand{\PACS}{68.37.Ps, 34.20.Cf, 68.35.Gy, 68.35.Ja}
%%%%%%%%%%%%%%%%%%%%%%%%%%%%%%%%%%%%%%%%%%%%%%%%%%%%%%%%%%%%%%%%%%%%%%%%%%%%%%
%% please enter (First) Author (et al.) and short version of the title here %%
%%%%%%%%%%%% must not exceed 80 characters in length together %%%%%%%%%%%%%%%%
%%%%%%%%%%%%%%%%%%%%%%%%%%%%%%%%%%%%%%%%%%%%%%%%%%%%%%%%%%%%%%%%%%%%%%%%%%%%%%
%\newcommand{\shorttitle}
%{F. J. Giessibl et al., Imaging atomic orbitals by Atomic Force Microscopy}
%% sets the header on oddpage
%%%%%%%%%%%%%%%%%%%%%%%%%%%%%%%%%%%%%%%%%%%%%%%%%%%%%%%%%%%%%%%%%%%%%%%%%%%%%%
%%%%%%%%%%%%%%%%%%%%%%%% here comes the title group %%%%%%%%%%%%%%%%%%%%%%%%%%
%%%%%%%%%%%%%%%%%%%%%%%%%%%%%%%%%%%%%%%%%%%%%%%%%%%%%%%%%%%%%%%%%%%%%%%%%%%%%%
\title{Imaging of atomic orbitals with the Atomic Force Microscope -- experiments and simulations}
%%%%%%%%%%%%%%%%%%%%%%%%%%%%%%%%%%%%%%%%%%%%%%%%%%%%%%%%%%%%%%%%%%%%%%%%%%%%%%
\author{F. J. Giessibl, H. Bielefeldt, S. Hembacher, and J. Mannhart}
\date{June 27, 2001} % HAB

%%%%%%%%%%%%%%%%%%%%%%%%%%%%%%%%%%%%%%%%%%%%%%%%%%%%%%%%%%%%%%%%%%%%%%%%%%%%%%
\address
  {\hspace*{0.5mm} Universit\"{a}t Augsburg, Institute of Physics,
  Electronic Correlations and Magnetism,\\
  \hspace*{0.5mm} Experimentalphysik VI, 86135 Augsburg, Germany \\
  \tt franz.giessibl@physik.uni-augsburg.de}

\maketitle
%%%%%%%%%%%%%%%%%%%%%%%%%%%%%%%%%%%%%%%%%%%%%%%%%%%%%%%%%%%%%%%%%%%%%%%%%%%%%%
%%%%%%%%%%%%%%%%%%%%%%%%%%%%%%%%%%%%%%%%%%%%%%%%%%%%%%%%%%%%%%%%%%%%%%%%%%%%%
\begin{abstract}
  Atomic force microscopy (AFM) is a mechanical profiling
technique that allows to image surfaces
with atomic resolution. Recent progress in reducing the noise of this
technique has led to a resolution level where
previously undetectable symmetries of the images of
single atoms are observed. These symmetries are related to the nature of the interatomic
forces.
The Si(111)-(7$\times$7)
surface is studied by AFM with various tips and AFM images are simulated
with chemical and electrostatic model forces. The
calculation of images from the tip-sample forces is explained in detail
and the implications of the imaging parameters are discussed.
Because the structure of the Si(111)-(7$\times$7) surface is known
very well, the shape of the adatom images is used to
determine the tip structure.
The observability of atomic orbitals by AFM and
scanning tunneling microscopy is discussed.

\end{abstract}
%%%%%%%%%%%%%%%%%%%%%%%%%%%%%%%%%%%%%%%%%%%%%%%%%%%%%%%%%%%%%%%%%%%%%%%%%%%%%

\section{Introduction}

In 1959, Schlier and Farnsworth reported their low-energy electron diffraction
(LEED) experiments on the surface of Si (111) \cite{Schlier1959}. After heating
the surface to 900\,${{}^\circ}$C, they discovered that the surface displays
additional scattering peaks in the LEED pattern. These additional peaks were caused by
a reconstruction of the Si surface where the new unit cell is 7$\times $7
as large as the bulk terminated structure.
Silicon condenses in the diamond structure with a cubic lattice constant
of $a_0=5.430$\,\AA{} at $T=0$\,K and $a_0=5.4355$\,\AA{} at $T=300$\,K \cite{Landolt1982}.
The unit vectors of the bulk terminated Si (111) surface
have a length of $v=a_0/\sqrt{2}=3.84$\,\AA{} and an angle of 60${{}^\circ}$, in the
reconstructed surface, the unit vectors have a length of $w=7 \times 3.84$\,\AA{}
$=26.88$\,\AA{} (see Fig. 1).
\parbox{17cm}{\begin{figure}[htbp]
  \centering
  \centerline{\epsfxsize=12cm{\epsfbox{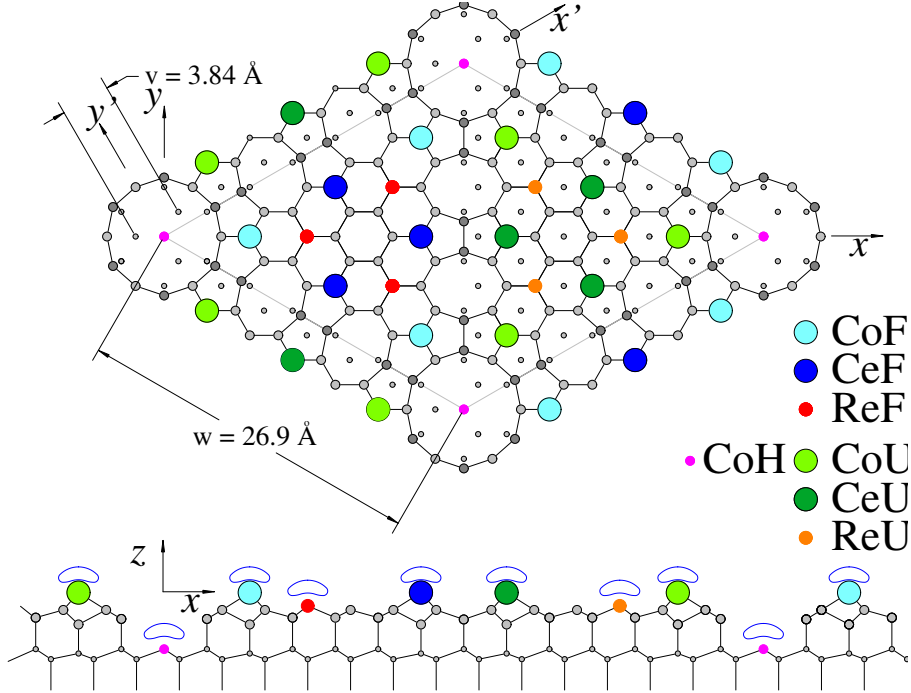}}}
  \caption{Top view and cross section ($y=0$) of the dimer-adatom-stacking fault
(DAS) model of Si (111)-(7$\times$7). The unreconstructed surface lattice has a
lattice constant of 3.84\,\AA.{} By forming the 7$\times$7
reconstruction, most of the dangling bonds of the surface atoms are
saturated by an adatom such that the total number of dangling bonds per
unit cell is reduced from 49 to 19. Twelve adatoms and a corner hole per
unit cell are the striking features of this reconstruction.
The adatoms fall into four symmetry
classes: corner faulted (CoF), center faulted (CeF), corner unfaulted (CoU),
 center unfaulted (CeU). The 19 remaining dangling bonds originate from
 the 12 adatoms, the 6 rest atoms (ReF, ReU) and the atom in the center of
 the corner hole (CoH).}\label{DAS}
\end{figure}}
The reconstruction not only affects the surface atom layer, but
the topmost four layers, and the new unit cell contains
approximately 200 atoms. Because of this remarkable size of the
unit cell, the determination of the atom positions has been a
tremendous challenge for surface scientists for more than two
decades. The determination of the positions of the surface atoms
by Binnig {\it et al.} with a scanning tunneling microscope (STM)
in 1982 \cite{Binnig1983} was instrumental for the discovery of
all the atomic positions within the surface unit cell. In 1985,
the now commonly adapted dimer-adatom-stacking-fault model (DAS)
was finally suggested by Takayanagi {\it et al.}
\cite{Takayanagi1985}. According to this model (see Fig. 1), six
adatoms are situated in each half of the unit cell. The adatoms
are bound by covalent bonds formed by $sp^{3}$ hybrid orbitals. In
the bulk, the hybrid orbitals of neighboring atoms overlap and
form an electron pair. At the surface, one of the four $sp^{3}$
hybrid orbitals is pointing perpendicular to the surface and forms
a dangling bond.

Because of its historic role in the establishment of the STM, the
Si (111)-(7$\times $7) surface has been considered as a touchstone
for the feasibility of the atomic force microscope (AFM)
\cite{Binnig1986} as a tool for surface science. However, because
of the reactivity of Si (111)-(7$\times $7), AFM experiments have
proven to be difficult in the quasistatic AFM mode, and chemical
bonding between tip and sample has hampered atomic resolution
\cite{Howald1995}. In 1994, Si (111)-(7$\times $7) was imaged by
AFM for the first time \cite{Giessibl1995,Kitamura1995}. AFM is
now becoming an increasingly powerful tool for surface science, as
true atomic resolution is achieved routinely
\cite{NCAFM1998,NCAFM1999,NCAFM2000}. Surfaces are imaged with
atomic resolution by bringing a probe into close proximity and
sensing the interaction while scanning the surface. It has been
shown that a dynamic mode of imaging is required for resolving
reactive surfaces like Si (111) in ultra-high vacuum where
chemical bonding between tip and sample can occur \cite
{Giessibl1995,Kitamura1995}. Frequency-Modulation AFM (FM-AFM)
\cite {Albrecht1991} has proven to be a practical method for fast
imaging in vacuum and is now the commonly used technique.
Recently, it was proposed \cite{Giessibl1999b} that the noise
decreases by a factor of approximately 10 by optimizing the
stiffness of the cantilever and the oscillation amplitude (see
section \ref{experimental}). Consequently, unprecedented
resolution has been achieved with these operating parameters.
Especially, deviations from the so far hemispherical apparent
shape of the individual atoms have been observed, and these
deviations have been attributed to the symmetry of the chemical
bonds between tip and sample.

The first AFM images of Si (111)-(7$\times $7) \cite{Giessibl1995,Kitamura1995}
looked similar to the STM images of the empty states. The bond between the front
atom of the tip and the dangling bond on top of the adatoms is thought to be
responsible for the experimentally observed atomic contrast in the FM-AFM
images \cite{Perez1997,Perez1998}. While the atom in the center of the
cornerholes is $4.45$\,\AA{} below the adatoms surrounding the
cornerhole \cite{Brommer1993}, in typical STM
experiments a depth of $2$\,\AA{} is observed. The discrepancy between the
nominal depth of the cornerholes and the STM data is probably caused by the
finite tip radius of the STM tips. The tip is not sharp enough to allow a
penetration of the front atom all the way to the bottom of the cornerhole.
In the first AFM
experiment, the observed depth of the cornerhole was only
$\approx 0.8$\,\AA{} \cite{Giessibl1995}. This
even smaller depth is probably due to a significant contribution of long-range
forces. It is shown here, that the contribution of long-range forces is reduced
by operating the microscope with very small oscillation amplitudes and a
cornerhole depth of $\approx 2.5$\,\AA{} is obtained.

In the present work, the Si(111)-(7$\times $7) surface is investigated using FM-AFM
with optimised resolution.
The structure of the Si(111)-(7$\times$7) surface
is known very well through a large number of experimental and theoretical studies.
Because of its pronounced features, such as the
deep cornerholes and the widely spaced adatoms with a well defined bonding
characteristic, the Si(111)-(7$\times$7)
surface is a perfect sample for studying the atomic and subatomic structure of the tip.
The spatial resolution of the
AFM has advanced sufficiently such that clear deviations of an $s$- or $p_z$- type
tip have been observed \cite{Giessibl2000c}.
Therefore, in contrast to previous studies, tip and
sample switch roles now and the well defined surface is used to probe the tip, where
much less information about its structure and composition is available.

\section{Experimental setup}\label{experimental}

The data was taken with a modified commercial scanning
tunneling microscope \cite{ThermoMicroscopes,Giessibl1994a} which has been
outfitted with a force sensor based on a quartz tuning fork
(\lq\lq qPlus-sensor\rq\rq, \cite{Giessibl2000}). The tip of the force sensor
is either an etched tungsten tip such as known from STM or a silicon crystallite.
In FM-AFM, a
cantilever beam holding a sharp tip with an
eigenfrequency $f_{0}$, spring constant $k$ and quality factor $Q$ is subject to
controlled positive feedback such that it oscillates with a constant
amplitude $A$. When this cantilever is brought close to a sample, its
frequency changes from $f_{0}$ to $f=f_{0}+\Delta f$. This frequency change $%
\Delta f$ is used to create an image $z(x,y,\Delta f)$ by scanning the
cantilever in the $x-y$ plane and recording the corresponding $z-$position
of the base of the cantilever $z_{b}$. A feedback mechanism adjusts $z_{b}$
such that $\Delta f$ stays constant. In the ``classic''{} mode of
operation, typical parameters are $\Delta f\approx -100$\thinspace Hz, $k\approx 20$
\thinspace N/m, $f_{0} \approx 200$\thinspace kHz, $A\approx 10$\thinspace
 nm and $Q\approx 10^{5}$ - see Table 1 in Ref.
\cite{Giessibl1999b} for an overview. Equation \ref{z_noise} shows that the vertical resolution
in FM-AFM can be improved by reducing the amplitude $A$ to the
\AA-range \cite{Giessibl1999b}. For tip-sample forces
$F_{ts}=F_0\exp(-z/\lambda)$ where
$z$ is the tip-sample distance, the vertical noise $\delta z$ is a
function of the amplitude $A$, the force range $\lambda$, the detection bandwidth $B$ and the temperature $T$:
\begin{equation}
\delta z \propto  \frac{(1+\sqrt{\frac{\pi }{2}}(\frac{A}{\lambda })^{3/2})}{A}
\sqrt{TB}.
\label{z_noise}
\end{equation}
Minimal noise results when $A \approx \lambda$ and for a small product of temperature and bandwidth.
The bandwidth $B$ is approximately equal to the number of pixels per second
which can be recorded. Operating at a small bandwidth implies a small scanning speed and
long image acquisition times, and thermal drift can become a problem. When operating at low temperatures,
thermal drift is usually not present and imaging at very small scanning speeds is feasible.
Operation with very small amplitudes is not possible with conventional cantilevers. In
order to avoid instabilities like \lq\lq jump-to-contact\rq\rq and to minimize perturbations of the
oscillation by tip-sample dissipation, it helps to use cantilevers with $k\approx 1000$\,N/m and $Q\approx 1000$
\cite{Giessibl2000Habil}.
A secondary benefit of using small amplitudes is an enhanced sensitivity to short-range forces and
a reduced sensitivity to long-range forces (see Fig. \ref{convolution_deltaf}).
\parbox{17cm}{
\begin{figure}[htbp]
  \centering
  \centerline{\epsfxsize=12cm{\epsfbox{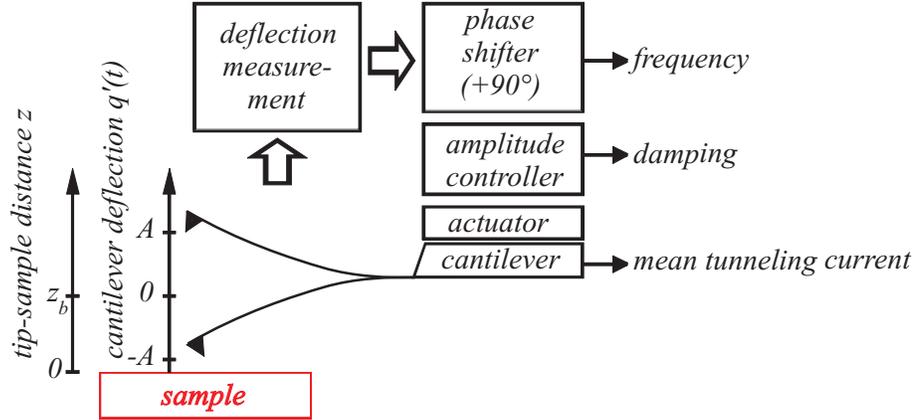}}}
  \caption{Operating principle of a frequency modulation AFM.
  A cantilever is mounted onto an actuator, which is used
  to excite an oscillation at an adjustable fixed amplitude $A$. When forces act between
  tip and sample, the oscillation frequency changes. Additional observables are the driving
  signal which is required to operate the cantilever at a given amplitude. If sample and
  cantilever tip are electrically conductive, a mean tunneling current can also be measured.}\label{FMmode}
\end{figure}}
The oscillator electronics
which controls the force sensor is designed with an emphasis on low noise such
that oscillation amplitudes in the \AA-range can be realized. The commercial
phase-locked-loop (PLL)
frequency detector is quartz stabilized \cite{Nanosurf}.
In the vacuum chamber, pumped with an ion and
titanium sublimation pump, a base pressure of $5\times 10^{-11}$ mbar is
reached. As a sample, a $10\times 13$\, mm$^{2}$ piece of a p-doped (B) silicon (111)
wafer with a resistivity of 9\,$\Omega$cm is used and the
7$\times$7 - reconstruction is obtained by heating to $1300{%
^{\circ }}$C for 30\thinspace s with an electron beam heater.

Typically, the experiment
is started by operating the microscope in the STM mode, i.e. the feedback is
controlled by the tunneling current, because the tip can be cleaned and
conditioned by applying voltage pulses. The frequency shift is recorded in parallel with the
topography.
The tunneling
current is collected at the sample and a typical setpoint of the time-averaged
tunneling current is $100$\,pA.
The surface is scanned in search for a clean spot which displays
large areas of reconstructed silicon. Also, controlled collisions between
tip and sample are performed in order to shape the tip such that good
STM images are obtained. The frequency shift which occurs during STM imaging is used as
a reference before switching into the AFM mode. Then, the tip is withdrawn and the
feedback is switched to frequency shift control. The setpoint of the
frequency shift is slowly decreased while the tip is scanning. While
carefully decreasing the setpoint of the frequency shift, the corrugation of the image
is increasing until an optimal value is reached.

One complication of AFM versus STM is that unlike in STM, the imaging signal
in AFM is not monotonic with respect to distance. The tunneling current increases monotonically with
decreasing tip-sample distance. However, the tip-sample force is in general
attractive for large tip-sample distance and turns repulsive at a distance of
the order of the equilibrium distance in the bulk crystal. Consequently, the
frequency shift is also non-monotonic.

Figure \ref{gamma_min} shows
a schematic curve of a short-range tip sample interaction curve and the
corresponding observable in FM-AFM - the normalized frequency shift
$\gamma _{lA}(z)$- curve (see Eq. \ref{gamma_def}).

\parbox{17cm}{
\begin{figure}[htbp]
  \centering
  \centerline{\epsfxsize=8.4cm{\epsfbox{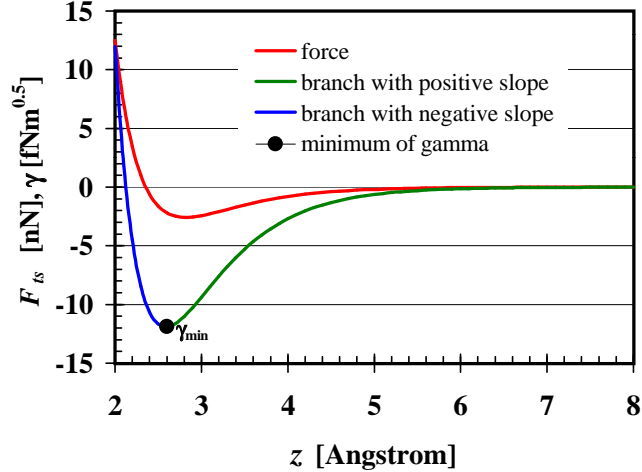}}}
  \caption{Short-range force modelled by a Morse type force
  $F_{ts}(z) = -2E_{bond}\kappa[\exp(-\kappa(z-\sigma))+\exp(-2\kappa(z-\sigma))]$
  with $E_{bond}=2.15$\,eV, $\sigma=2.35$\,\AA{} and $\kappa=1.5$\,\AA$^{-1}$ and
  corresponding normalized frequency shift $\gamma (z)$.}\label{gamma_min}
\end{figure}}
Because the feedback in an FM-AFM uses the frequency shift as the
error signal, stable feedback is only possible on a monotonic
branch of the frequency shift curve. When operating on the branch
with positive slope, the feedback system withdraws the cantilever
from the surface when the actual frequency shift is smaller (i.e.
more negative) than the setpoint and vice versa. When operating at
the branch with negative slope, the sign of the control output has
to be reversed for obtaining proper feedback operation. The two
branches meet at $z=z_{\gamma\,min}$ with
$\gamma(z_{\gamma\,min})=\gamma_{min}$. L\"{u}thi {\it et al.}
have identified the distance range where atomic resolution is
possible and found that atomic resolution is obtained for $z$
close to $z_{\gamma\,min}$ \cite{Luethi1996}. For the frequency
shift curve shown in Fig. \ref{gamma_min}, stable feedback on the
branch with positive slope is only possible for $z\geq 2.6$\,\AA.
For accessing a smaller distance range, an additional long-range
force can be added by applying an electric field. Hence,
$z_{\gamma\,min}$ is shifted to a smaller distance and tip and
sample can get closer while the feedback still operates correctly.
This is important for imaging atomic orbitals as evident from Fig.
\ref{sto}.

\section{Observables in FM-AFM and physical interpretation of images}

When the gradient of the tip sample forces $k_{ts}$ is constant for the
whole $z$ range of the tip motion, the frequency shift is given by \cite
{Albrecht1991}
\begin{equation}
\Delta f(z_{b})=f_{0}\frac{1}{2k}k_{ts}(z_{b})  \label{df small A}
\end{equation}
where $z_{b}$ is the mean $z-$position of the cantilever tip. If $k_{ts}(z)$
is not constant for the $z-$range covered by the oscillating cantilever $%
z_{b}-A<z<z_{b}+A$, $\Delta f$ can be calculated by first order perturbation
theory using the Hamilton-Jacobi approach \cite{Giessibl1997}
\begin{equation}
\Delta f(z_{b})=-\frac{f_{0}^{2}}{kA^{2}}\left\langle F_{ts}(z_{b}+q^{\prime
})q^{\prime }\right\rangle  \label{df1}
\end{equation}
where $z_{b}$ is the vertical base position of the cantilever as shown in
Fig. \ref{FMmode}. The precision of $\Delta f$ obtained by first order perturbation
theory is determined by the ratio between the magnitude of the perturbation
(i.e. the tip sample potential $V_{ts}$) and the energy $E$ of the
oscillating CL. This ratio is in the order of $10^{-3}$ both in classic
FM-AFM with soft cantilevers and large amplitudes ($k\approx 20\,$N/m, $%
A\approx 10\,$nm) as well as in non-contact AFM operated with very stiff cantilevers
and small amplitudes ($k\approx 2000\,$N/m, $A\approx 1\,$nm). In both
cases, the energy $E=kA^{2}/2$ of the cantilever is $\approx 6\,$keV while
the magnitude of $V_{ts}$ does not exceed a few eV, at least for the cases
where atomic resolution is desired and the interaction between tip and
sample is dominated by the front atom of the tip.

Substituting $q^{\prime }=A\cos (2\pi f_{0}t)$, $z=z_{b}-A$ yields
\begin{equation}
\Delta f(z)=\frac{f_{0}}{\sqrt{2}\pi kA^{3/2}}\int_{0}^{2A}\frac{%
F_{ts}(z+z^{\prime })}{\sqrt{z^{\prime }}}\frac{1-z^{\prime }/A}{\sqrt{%
1-z^{\prime }/2A}}dz^{\prime }.  \label{deltafF}
\end{equation}
Integration by parts yields an expression which is closely related to Eq.
\ref{df small A} \cite{Giessibl2001}:
\begin{equation}
\Delta f(z)=\frac{f_{0}}{2k}\frac{2}{\pi A^{2}}\int_{0}^{2A}k_{ts}(z+z^{%
\prime })\sqrt{2Az^{\prime }-z^{\prime 2}}\,dz^{\prime }.  \label{deltaFkts}
\end{equation}
The frequency shift is proportional to an average tip-sample force gradient,
which is calculated by weighing the force gradient with a semi-circle with
radius $A$ (Fig. \ref{convolution_deltaf}).

\parbox{17cm}{
\begin{figure}[htbp]
  \centering
  \centerline{\epsfclipon\epsfxsize=8.4cm {\epsfbox{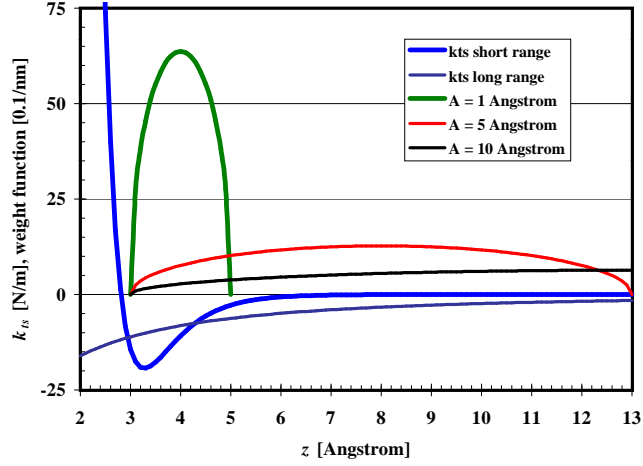}}}
  \caption{Tip-sample force gradient $k_{ts}$ and weight function for the calculation
  of the frequency shift.}\label{convolution_deltaf}
\end{figure}}
Thus, the use of small amplitudes increases the relative
contribution of short-range forces and reduces the contribution of
long-range forces to the frequency shift.

The frequency shift is a function of four parameters: $k,A,f_{0}$, and $z$.
It is useful to exploit the scaling properties of $\Delta f$ in order to be
able to scale calculated frequency shifts to a variety of experimental
situations. The frequency shift is strictly proportional to $f_{0}$ and $1/k$%
, and if $A$ is large compared to the range of the forces (this condition is
usually fulfilled, see below), it scales as $A^{-3/2}$. Therefore, it is
useful to define a ``normalized frequency shift''
\begin{equation}
\gamma (z,A)=\frac{\Delta f(z,f_{0},k,A)}{f_{0}}kA^{3/2}.\label{gamma_def}
\end{equation}
Explicitly, we find
\begin{equation}
\gamma (z,A)=\frac{1}{\sqrt{2}\pi }\int_{0}^{2A}\frac{F_{ts}(z+z^{\prime })}{%
\sqrt{z^{\prime }}}\frac{1-z^{\prime }/A}{\sqrt{1-z^{\prime }/2A}}dz^{\prime
}.  \label{gammaFts}
\end{equation}
Again, we can integrate by parts and we find that $\gamma$ is also given
by the convolution of $k_{ts}$ with a weight function:
\begin{equation}
\gamma (z,A)=\frac{1}{\sqrt{2}\pi }\int_{0}^{2A}k_{ts}(z+z^{\prime })\sqrt{%
z^{\prime }-\frac{z^{\prime 2}}{2A}}dz^{\prime }.  \label{gammakts}
\end{equation}
\parbox{17cm}{
\begin{figure}[htbp]
  \centering
  \centerline{\epsfxsize=8.4cm{\epsfbox{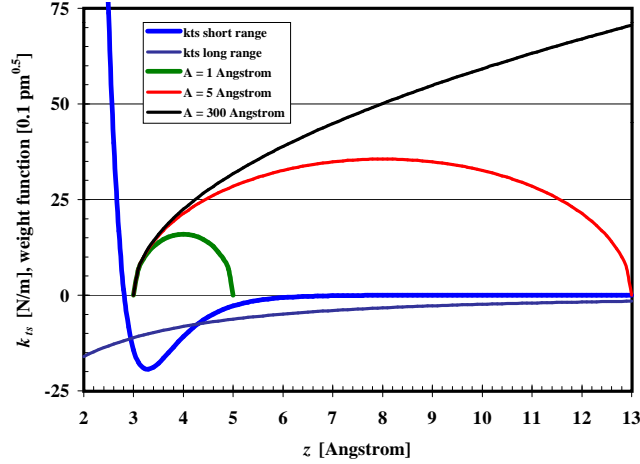}}}
  \caption{Tip-sample force gradient $k_{ts}$ and weight function for the calculation
  of the normalized frequency shift $\gamma$. If the amplitude $A$ is large compared
  to the range of $k_{ts}$, the value of $\gamma$ saturates and is no longer
  a function of $A$.}\label{convolution_gamma}
\end{figure}}
If the amplitude $A$ is large compared to the range of the tip-sample
forces, the normalized frequency shift $\gamma (z,A)$ asymptotically reaches
a large amplitude limit $\gamma _{lA}(z)$. This is explicitly shown for
inverse power, power- and exponential forces in Ref. \cite{Giessibl2000b}.
For Stillinger-Weber forces (see below), the deviation between
$\gamma (z,A)$ and $\gamma _{lA}(z)$ is less than 3\,\% for $A\geq 8$\,\AA.
For large amplitudes, we find
\begin{equation}
\gamma _{lA}(z)=\frac{1}{\sqrt{2}\pi }\int_{0}^{\infty }\frac{%
F_{ts}(z+z^{\prime })}{\sqrt{z^{\prime }}}dz^{\prime }
=\frac{1}{\sqrt{2}\pi }%
\int_{0}^{\infty }k_{ts}(z+z^{\prime })\sqrt{z^{\prime }}dz^{\prime }.
\label{gammalA}
\end{equation}
Experimental FM-AFM images can be created in the topographic mode, where the
frequency shift is kept constant and the $z$- position is adjusted accordingly
as a function of $x$ and $y$ and the \lq\lq constant height\rq\rq{} mode
where the frequency shift is recorded for constant $z$ as a function of
$x$ and $y$.
Thus, topographic FM-AFM images correspond to a three-dimensional map of
$z(x,y,\gamma)$, and constant height images correspond to a
three-dimensional map of $\gamma(x,y,z,A)$.
Because the frequency shift of the cantilever is only a function of the
$z-$ component of the tip-sample force, $\gamma(x,y,z,A)$ is simply given by:

\begin{equation}
\gamma (x,y,z,A)=\frac{1}{\sqrt{2}\pi }\int_{0}^{2A}\frac{F_{ts}(x,y,z+z^{\prime })}{%
\sqrt{z^{\prime }}}\frac{1-z^{\prime }/A}{\sqrt{1-z^{\prime }/2A}}dz^{\prime
}.  \label{gammaFts3d}
\end{equation}

It is noted, that Eq. \ref{gammaFts3d} implies that
the cantilever vibrates exactly parallel to
the surface normal vector of the sample. Conventional cantilevers have a tip which is
only a few micrometers high, therefore they have to be tilted by an angle
$\alpha \approx $10\,$^{\circ}$
to make sure that among all the parts
of the cantilever and cantilever holder, the tip is the closest part
to the sample. If the cantilever is tilted around the
$y-$ axis, $\gamma$ is given by
\begin{equation}
\gamma (x,y,z,A)=\frac{1}{\sqrt{2}\pi }
\int_{0}^{2A}\frac{F_{ts}(x+q^{\prime }\sin\alpha ,y,z+q^{\prime }\cos\alpha)}{%
\sqrt{q^{\prime }}}\frac{1-q^{\prime }/A}{\sqrt{1-q^{\prime }/2A}}dq^{\prime
}.  \label{gammaFts3dtilted}
\end{equation}
The qPlus sensor does not need to be tilted because it can be outfitted with large tips
(see Fig. 1 in \cite{Giessibl2000}) and in the simulations presented here, $\alpha = 0$ is assumed.

D\"{u}rig has devised an expanded approach for the calculation of the frequency shift which
also encompasses the damping effects \cite{Durig2000}.

\section{Calculation of tip-sample forces}

The first step for the simulation of images is the calculation of
the tip sample forces $F_{ts}$. One approach for this task is the
use of the density functional theory and related methods, such as
shown by Perez {\it et al.} \cite{Perez1998}. However, as
emphasized by Bazant {\it et al.}  \cite{Bazant1997} the use of
empirical force models still has its merits, because {\it ab
initio} methods require much more computing power and programming
effort than the use of empirical force models. Here, we estimate
the force between the tip of a cantilever and a flat sample
$F_{ts}(x,y,z)$ by a long-range component that is only a function
of vertical tip-sample distance $z$ and a short-range force that
also depends on the lateral coordinates $x$ and $y$:
\begin{equation}
F_{ts}(x,y,z) = F_{lr}(z)+F_{sr}(x,y,z).
\label{Ftsgen}
\end{equation}

\subsection{Van-der-Waals forces}
In principle, the van-der-Waals (vdW) force is a short-range
force, characterized by the $1/r^6$ vdW potential. However, the
vdW force between two individual atoms is weak compared to the
chemical and electrostatic short-range forces. Nevertheless, the
total vdW force between a typical AFM tip and a flat sample can be
substantial. When assuming that the vdW force is additive, the sum
over the individual pair contributions can be replaced by an
integration over the volume of the tip and sample (Hamaker
integration) \cite{Hamaker1937,Israelachvili1991}. For a spherical
tip with radius $R$, the vdW force is given by
\begin{equation}
F_{vdW}(\zeta)=-\frac{A_H R}{6\zeta^2}=\frac{C_{vdW}}{\zeta^2}
\end{equation}
where $A_H$ is the Hamaker constant and $\zeta$ is the distance
between the center of the front atom of the tip and the plane connecting
the centers of the top layer of the sample atoms.
For a Si tip and a Si sample, $A_H=0.186$\,aJ \cite{Senden1995}.
The corresponding normalized frequency shift is \cite{Holscher1999,Giessibl2000b}
\begin{equation}
\gamma_{vdW}(\zeta,A) =  \frac{1}{ (2+\zeta/A)^{3/2} } \frac{C_{vdW}}{\zeta^{3/2}}.
\end{equation}

\subsection{Chemical bonding forces}
If tip and sample consist of silicon, the Stillinger-Weber (SW)
potential \cite {StillingerWeber1985} can be used to model the
chemical interaction. As noted by Bazant {\it et al.}
\cite{Bazant1997}, the SW potential is a fairly good model for the
mechanical properties of silicon. The SW
potential has been used before for the simulation of AFM images of Si(001)-(2%
$\times $1) in the quasistatic mode by Abraham {\it et al.} \cite
{Abraham1988}.

The SW potential necessarily contains nearest and next nearest
neighbor interactions. Unlike solids with a face centered cubic or body
centered cubic lattice structure, solids which crystallize in the diamond
structure are unstable when only next-neighbor interactions are taken into
account. The nearest neighbor contribution of the SW potential is
\begin{eqnarray}
V_{n}(r)&=&E_{bond}A\left[ B(\frac{r}{\sigma ^{\prime }})^{-p}-(\frac{r}{%
\sigma ^{\prime }})^{-q}\right] \exp\left({\frac{1}{r/\sigma ^{\prime }-a}}\right)
\,\,{\rm  for }\,\,r<a\sigma ^{\prime },\nonumber\\
&=0&\,\,{\rm for}\,\,r>a\sigma ^{\prime }.
\end{eqnarray}
\label{Vn}
The next nearest neighbor contribution is
\begin{equation}
V_{nn}({\rm\bf r}_{i},{\rm\bf r}_{j},{\rm\bf r}_{k})=E_{bond}[
h(r_{ij},r_{ik},\theta _{jik})
+h(r_{ji},r_{jk},\theta_{ijk})+h(r_{ki},r_{kj},\theta _{ikj})]
\end{equation}
\label{Vnn} with

\begin{eqnarray}
h(r_{ij},r_{ik},\theta _{jik})&=&
\lambda \exp\left({ \frac{\gamma}{r_{ij}/\sigma^{\prime }-a}
+\frac{\gamma}{r_{ik}/\sigma ^{\prime }-a}}\right) (\cos \theta _{jik}+\frac{1}{3})^{2}
\,\,{\rm  for }\,\,r_{ij,ik}<a\sigma ^{\prime }\nonumber\\
&=0& \,\,{\rm  for }\,\,r_{ij,ik}>a\sigma ^{\prime }.
\end{eqnarray}
Stillinger and Weber found optimal agreement with experimental
data for the following parameters:
\begin{center}
\begin{tabular}{ccc}
$A=7.049556277   $ &   $p=4$ &   $\gamma =1.20$ \\
$B=0.6022245584  $ &   $q=0$ &   $\lambda =21.0$ \\
$E_{bond}=0.34723$ aJ & $a=1.8$ &   $\sigma ^{\prime }=2.0951$ \AA
\end{tabular}
\end{center}

The equilibrium distance $\sigma $ is related to $\sigma ^{\prime }$ by $%
\sigma =2^{1/6}\sigma ^{\prime }$. The potential is constructed in such a
way to ensure that $V_{n}$ and $V_{nn}$ and all their derivatives with
respect to distance vanish for $r>a\sigma ^{\prime }=3.7718$\thinspace \AA.
The diamond structure is favoured by the SW potential because of the factor $%
(\cos \theta +\frac{1}{3})^{2}$ -- this factor is zero when
$\theta $ equals the tetrahedon bond angle of $\theta
_{t}=109.47^{\circ }$. The SW potential leads to a maximal
chemical contribution to the normalized frequency shift of $\gamma
\approx - 10\,$fNm$^{1/2}$. This has been verified with great
precision in an experiment where the electrostatic interaction was
carefully minimized by Lantz {\it et al.}
\cite{Lantz2000,Lantz2001}. Recently, Laschinger has performed
tight binding calculations for a silicon tip and a silicon sample
\cite{Laschinger2001} and has also found a double peak in $\gamma$
with a magnitude of $\gamma \approx - 10$\,fNm$^{1/2}$. However,
the distance at which this double peak occurs is much higher in
the tight binding calculation than in the SW result.

\subsection{Electrostatic forces}
The electrostatic force for a spherical tip with radius $R$ at a
differential voltage $U$ with respect to a flat surface at distance $\zeta$ is
given by \cite{Olsson1998}
\begin{equation}
F_{lr\,est}(\zeta)=-U^{2}\pi\epsilon_{0}\frac{R}{\zeta}=\frac{C_{est}}{\zeta},\label{est_lr}
\end{equation}
for $R \gg \zeta$, where $U$ is the bias voltage between tip and sample.
This formula is only correct for $\zeta \gg \sigma$
where $\sigma$ is the nearest neighbor distance in the bulk material of tip and
sample.
The corresponding normalized frequency shift is \cite{Holscher1999,Giessibl2000b}
\begin{equation}
\gamma_{est}(\zeta,A)=\frac{C_{est}}{\sqrt{2\zeta}}(\frac{1+\zeta/A}{\sqrt{1+\zeta/(2A)}}-\sqrt{\frac{2\zeta}{A}}).
\end{equation}
For a very small tip-sample distance, there will also be
short-range electrostatic forces. The electrostatic potential is very large close to
the nuclei of the atoms, and because the electrons shield this
field, the electrostatic potential decays exponentially (see Eq. \ref{fnl}).
If a bias voltage $U$ is applied between tip and sample, surface charges
will result and the electrostatic field has a component which decreases exponentially
with the surface separation with a decay length $\lambda =
\sigma/(2\pi)$ \cite{Feynman1977}. If a bias $U$ is applied between tip and sample,
we can roughly estimate the charge that is induced in the tip and sample atom
with the image charge method. We treat both tip and sample atom as a conducting
sphere with radius $\rho$ at a distance between their centers of $r$.
What is the charge induced on the spheres if we vary $r$
we keep the voltage differential $U$ constant?
\parbox{17cm}{
\begin{figure}[htbp]
  \centering
  \centerline{\epsfxsize=7.2cm{\epsfbox{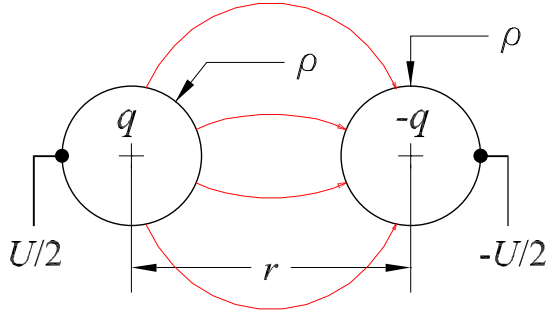}}}
  \caption{Electrostatic interactions of two spheres with radius $\rho$ and
  distance $r$ at a voltage differential $U$.}\label{spheres}
\end{figure}}
The voltage differential at the two spheres leads to an induced charge in
the spheres which is approximately given by
\begin{equation}
q(r)=2\pi\epsilon_{0}U\rho(1+\frac{\rho}{r-2\rho}).
\end{equation}
This follows from an electrostatic analogy: the equipotential surface of a point
charge is a sphere, and for two charges separated by $r$, the equipotential
surfaces again are approximately spherical. If we model the front atom of
the tip and the sample atom closest to the tip by a sphere with electrostatic
potential $U/2$ and $-U/2$ respectively, we can estimate the potential energy by
\begin{equation}
V_{sr\,estatic}(r)=-\pi\epsilon_{0}U^2\rho^2(1+\frac{\rho}{r-2\rho})^2\frac{1}{r}.\label{estatic_sr}
\end{equation}
Hence, the force is given by $F_{sr\,estatic}=-\partial V_{sr\,estatic}/\partial z$ with
$r=\sqrt{x^2+y^2+z^2}$.
The total electrostatic force is then calculated by summing a long-range electrostatic
force of a tip with a tip radius $R$ according to Eq. \ref{est_lr} and
a short range force given by the short range potential after Eq. \ref{estatic_sr}.
This model is shown in Fig. \ref{est_ts}.
\parbox{17cm}{
\begin{figure}[htbp]
  \centering
  \centerline{\epsfxsize=12cm{\epsfbox{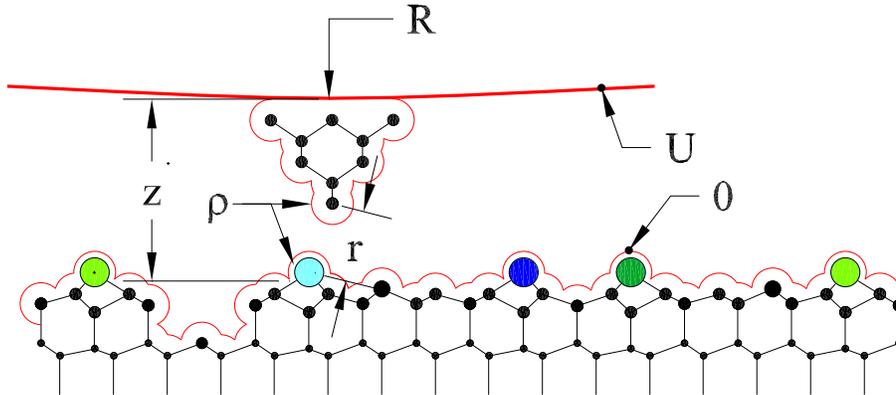}}}
  \caption{Electrostatic interactions of a macroscopic tip with a radius $R$
  plus a front atom with radius $\rho$ at a voltage differential $U$.}\label{est_ts}
\end{figure}}
We can thus estimate the electrostatic force by a long-range contribution
caused by the interaction of a macroscopic tip with radius $R$ with a flat surface
plus a short-range contribution describing the interaction of the front atom
with the sample atom closest to it.

The model of atoms as perfectly conducting spheres is certainly only a rough approximation. When getting
down to atomic length scales, the induced charges will not spread over the surface of
a sphere with the atomic radius. Instead,
the induced charges will modify the filling of the electronic states at tip and sample
at the Fermi energy, i.e. the valence orbitals. Therefore, we first take a look at the
charge density of the valence states in silicon.
The electronic wave functions the valence shell
of atoms can be approximately constructed with the use of the Slater-type orbital
model \cite{McWeeny1991}. The radial functions of the wave function are given by
\begin{equation}
f_{nl}(r)=\frac {N_{nl}}{a_B^{3/2}}(\frac{r}{a_B})^{n-1}\exp\left( -\frac {Z_e r}{na_B}\right).
\label{fnl}
\end{equation}
where $n$ is the main quantum number, $N_{nl}$ is a normalization constant,
$a_B = 0.529$\,\AA{} is Bohr's radius and
$Z_e=Z-s$ where $Z$ is the nuclear charge and $s$ is a screening constant. For the
valence shell of silicon, $n=3$,
$Z=14$ and $s=9.85$ according to the procedure described in Ref. \cite{McWeeny1991}.
The normalization constants are given by
\begin{equation}
N_{30}=\sqrt\frac{2}{45\pi}(\frac {Z_e}{n})^{7/2} \,\,{\rm  and
}\,\, N_{31}=\sqrt\frac{2}{15\pi}(\frac {Z_e}{n})^{7/2}.
\end{equation}
The wave functions corresponding to the four $3sp^3$ orbitals are constructed by
\begin{eqnarray}
&&|\psi_1(x,y,z)\rangle =  \frac{1}{2}[f_{30}(r)+f_{31}(r)(+x/r+y/r+z/r)]\nonumber\\
&&|\psi_2(x,y,z)\rangle =  \frac{1}{2}[f_{30}(r)+f_{31}(r)(-x/r-y/r+z/r)]\nonumber\\
&&|\psi_3(x,y,z)\rangle =  \frac{1}{2}[f_{30}(r)+f_{31}(r)(+x/r-y/r-z/r)]\nonumber\\
&&|\psi_4(x,y,z)\rangle =  \frac{1}{2}[f_{30}(r)+f_{31}(r)(-x/r+y/r-z/r)]
\end{eqnarray}
\label{psi_sp3}
The maximal charge density occurs at a radius where $r^2f_{nl}^2(r)$ is maximal,
i.e. for
\begin{equation}
r_{max}=a_B\frac{n^2}{Z_e}.
\end{equation}
For silicon with $n=3$, $r_{max}=1.147$\,\AA{}, close to $\sigma/2=1.175$\,\AA{} where
$\sigma$ is the next neighbor distance in the bulk.
In our estimate of the electrostatic short-range interaction, we assume that the induced
charge is centered around the maximum of the charge distribution at $r_{max}=1.147$\,\AA{}
with a radius $\sigma/2$.
Figure \ref{sto} shows the calculated charge density.
\parbox{17cm}{
\begin{figure}[htbp]
  \centering
  \centerline{\epsfxsize=7.2cm{\epsfbox{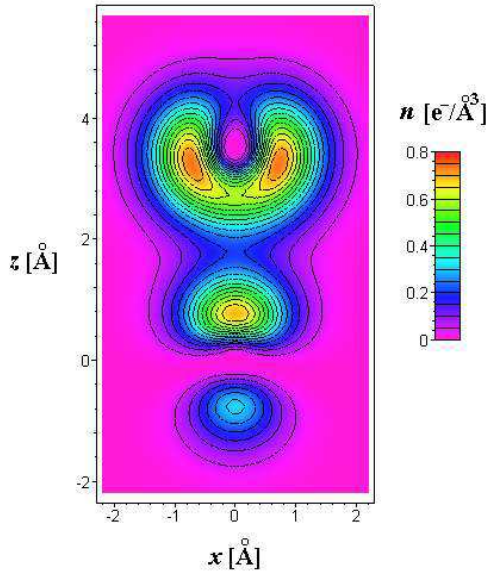}}}
  \caption{Charge density (in electrons per \AA$^3$) for one $3sp^3$ orbital
originating from an adatom located at $x=0$ and $z=0$ and two $3sp^3$ orbitals
originating from a tip atom located at $x=0$ and $z=3.5$\,\AA. The mean
density of valence electrons in bulk silicon is 0.2\,electrons per \AA$^3$.}\label{sto}
\end{figure}}
This figure shows, that in order to observe two separate maxima of the charge
density in an $xy-$ plane, the $z-$ distance of tip and sample atom needs to
be extremely small. Thus, imaging both dangling bonds separately is only
possible if the minimal
tip sample distance is of the order of the nearest neighbor
distance of the atoms in the bulk crystal. The calculation of the charge density
does not take the interaction of the adatom and the tip into account. Because
of Pauli's exclusion principle, the adatom
can only form a bond to one of the tip dangling bonds at the same time.

The long-range forces described with the models above vary strongly with
the macroscopic tip shape, and the short-range forces are a function of
the chemical identity of the front atom and its alignment to its bonding
partners in the tip. Once we have learned to prepare the AFM tip in a well
defined manner such that we know the atomic arrangement and chemical identity
of the tip apex, precise calculations of the tip-sample force are highly desirable.

\section{Experimental results and simulations}

Figure \ref{rtlltr} shows an experimental result of a
Si(111)-(7$\times$7) surface observed by FM-AFM. Every adatom
appears to have two peaks. This feature has been attributed to two
dangling bonds originating at the tip which image the single
dangling bond of the adatoms. While this image was recorded with a
tungsten tip, we think that the tungsten tip had a silicon
crystallite or cluster at its end. The notion, that tungsten tips
pick up silicon such that the front atom is made of silicon has
been put forth in STM experiments before
\cite{Demuth1988,Chen1991,Chen1993}.
\parbox{17cm}{
\begin{figure}[htbp]
  \centering
  \centerline{\epsfxsize=12cm{\epsfbox{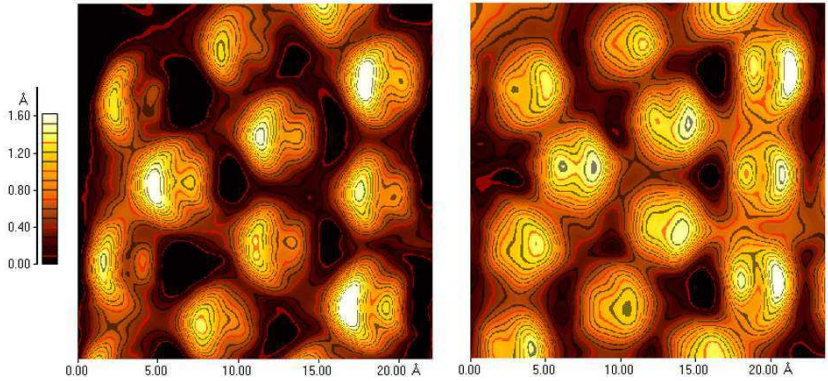}}}
  \caption{Topographic image of Si(111)-(7$\times$7) observed by FM-AFM,
  presumably imaged with a (001) oriented Si tip. Imaging parameters:
  $k=1800$\,N/m, $A=8$\,\AA, $f_0=16860$\,Hz, $\Delta f=-160$\,Hz, rms
  frequency error $\delta \Delta f=9$\,Hz, scanning speed $= 80$\,nm/s,
  sample voltage $U=1.6$\,V.}\label{rtlltr}
\end{figure}}
This interpretation of the data has been challenged recently \cite{Hug2001}. We
could show, that the feedback issues
proposed are not relevant in our experiment \cite{Giessibl2001b}. Further, the
asymmetry between the two scan directions is not due to a finite feedback speed,
but must be caused by elastic tip deformations such as illustrated in Figure \ref{tip_bending}.
\parbox{17cm}{
\begin{figure}[htbp]
  \centering
  \centerline{\epsfxsize=12cm{\epsfbox{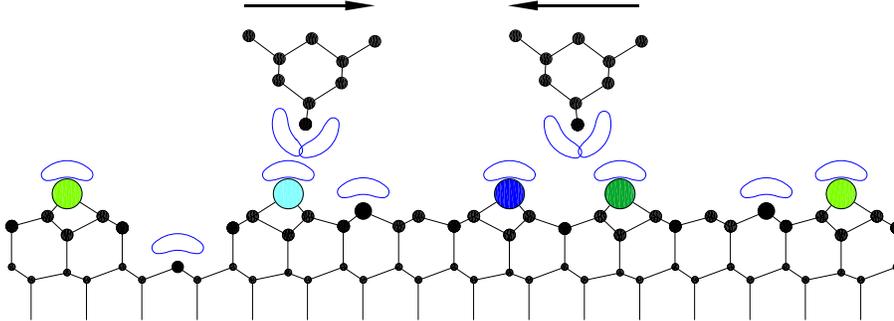}}}
  \caption{Tip bending effect: For a tip crystallite with the geometry
  proposed, the stiffness in scanning ($x$) direction is very small so it is likely that the
  tip apex bends opposite to the scan direction. This bending explains the reversal
  in the height of the two adatom peaks when the scanning direction reverses (Fig.
  9).}\label{tip_bending}
\end{figure}}
Figure \ref{s2peak} shows a magnified view of a single adatom. The
distance between the contour lines is 10\,pm, so the left peak is
about 60\,pm higher than the right peak and the depth between the
peaks is approximately 20\,pm.
\parbox{17cm}{
\begin{figure}[htbp]
  \centering
  \centerline{\epsfxsize=8.4cm{\epsfbox{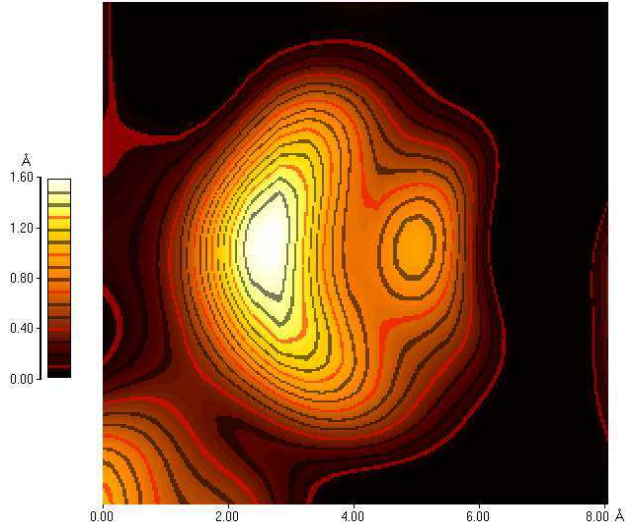}}}
  \caption{Topographic image of a single adatom on Si(111)-(7$\times$7) observed by FM-AFM,
  presumably imaged with a (001) oriented Si tip. The following imaging parameters were used: $k=1800$\,N/m,
$A=8$\,\AA, $f_0=16860$\,Hz, $\Delta f=-160$\,Hz, rms frequency
error $\delta \Delta f=9$\,Hz, scanning speed = 80\,nm/s, sample
voltage $U=1.6$\,V, scanning direction: left to
right.}\label{s2peak}
\end{figure}}
Using Eq. \ref{gammaFts3d}, we can calculate the expected image
once we know the tip-sample force. As we have noted in Ref.
\cite{Giessibl2000}, a qualitative picture can be simulated from a
long-range vdW force and a short range chemical force given by the
SW potential. Fig. \ref{gamvdW387} shows the result of that
calculation. The left image is simulated for the experimental
normalized frequency shift according to Figure \ref{s2peak} where
$\gamma = -160$\,Hz$/16860$\,Hz$\times
1800\,$N/m$\times(8\,$\AA$)^{3/2}=-387$\,fNm$^{1/2}$. The double
peaked adatom image can only occur if the long-range attractive
force is strong enough such that $\gamma_{min}$ occurs at a
distance where the charge density of the tip shows strong lateral
variations. The right image is simulated for $\gamma =
-4$\,fNm$^{1/2}$ - a typical value for weak vdW attractive forces
\protect\cite{Lantz2000,Lantz2001}. The left and the right images
are simulated with a Si tip which exposes two dangling bonds (see
Fig. \ref{100tip}). However, the two dangling bonds appear
separated only if the long-range attractive force is strong enough
to allow imaging at very small distances.
\parbox{17cm}{
\begin{figure}[htbp]
  \centering
  \centerline{\epsfxsize=12cm{\epsfbox{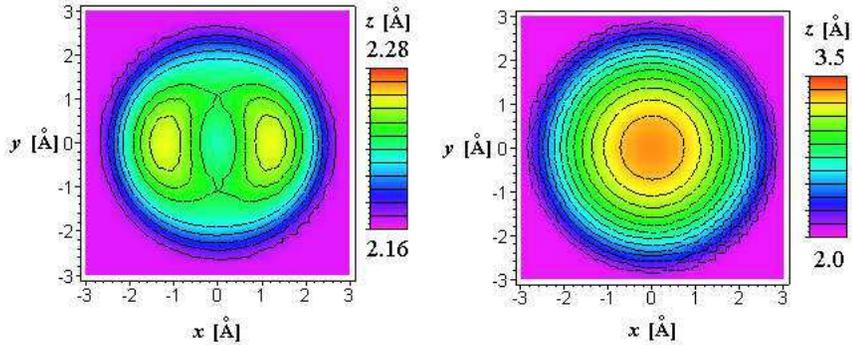}}}
  \caption{Simulation of a topographic FM-AFM image of a single adatom on
  Si(111)-(7$\times$7). Parameters (Left): $\gamma=-387$\,fNm$^{1/2}$, $A=8$\,\AA,
  short-range force: Stillinger Weber, tip geometry shown in Fig. 17,
  long-range force = $-3.66\times 10^{-26}$\,Jm/$(z+\Delta)^2$, $\Delta=5$\, \AA.
  Parameters (Right): $\gamma=-4$\,fNm$^{1/2}$, $A=8$\,\AA,
  short-range force: Stillinger Weber, tip geometry shown in Fig. 17,
  long-range force = $-3.66\times 10^{-28}$\,Jm/$(z+\Delta)^2$, $\Delta=5$\, \AA.
}\label{gamvdW387}
\end{figure}}
Several approximate procedures for obtaining $\gamma(x,y,z)$ which are
numerically much less expensive than the exact calculation after Eq. \ref{gammaFts3d}
have been proposed:
\begin{enumerate}
\item Ke {\it et al.} \cite{Ke1999} found that under certain conditions,
the geometric mean between force and energy
is roughly proportional to the frequency shift; \label{Kemethod}
\item Schwarz {\it et al.} \cite{Schwarz2000} propose to approximate $\gamma$ by
$\gamma \approx \frac{\sqrt{2}}{\pi}\frac{V_{ts}}{\sqrt{\lambda}}$ where $\lambda$ is
the range of the tip sample potential $V_{ts}$; \label{Schwarzmethod}
\item In an earlier publication \cite{Giessibl2000b} we proposed to
decompose the tip-sample interaction
into monotonic basic types $F_{ts}^i$ with an exponential-, power- or inverse power-
$z$-dependence and showed that
$\gamma \approx \frac{1}{\sqrt{2\pi}}\sum_{i}F_{ts}^i\sqrt{{V_{ts}^i}/{F_{ts}^i}}$.
\end{enumerate}
However, as pointed out by H\"{o}lscher {\it et al.}
\cite{Holscher2001}, the use of the approximative approaches leads
to inaccuracies for small tip-sample distances. Moreover, method
\ref{Kemethod} predicts that $\gamma(\xi) = 0$ for $V_{ts}(\xi)=0$
and $F_{ts}(\xi)=0$ and method \ref{Schwarzmethod} predicts that
$\gamma(\xi) = 0$ for $V_{ts}(\xi)=0$ which is incorrect. Because
we are simulating images at very small tip-sample distances we
refrain from using the approximate procedures and perform the
integration of $\gamma$ after Eq. \ref{gammaFts3d}. The
integration is implemented with Newton's method - a step width of
5\,pm has proven to provide sufficient accuracy.

Qualitatively, the calculated image is similar to the experimental image. However,
the adatom height in the calculated image is only 0.12\,\AA{} - a tenth of the
experimental height. This is because the electrostatic short
range interaction has been neglected. The relevance of the electrostatic short-range force is
confirmed by experimental results of atomic resolution on silicon with an applied
bias voltage where the normalized frequency shift was as low as $-180$\,fNm$^{1/2}$
\cite{Luethi1996} and even $-500$\,fNm$^{1/2}$ \cite{Guggisberg2000}.

In Figure \ref{gamvdW+est387}, we have also taken the short-range
electrostatic force into account.

\parbox{17cm}{
\begin{figure}[htbp]
  \centering
  \centerline{\epsfxsize=7.2cm{\epsfbox{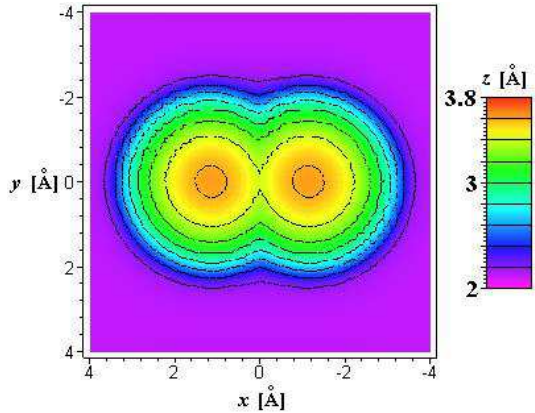}}}
  \caption{Simulation of a topographic FM-AFM image of a single adatom on
  Si(111)-(7$\times$7). Parameters: $\gamma=-387$\,fNm$^{1/2}$, $A=8$\,\AA, $U=1.6$\,V,
  short-range electrostatic force after Eq. 21
  tip radius $R=1000$\,\AA, Hamaker constant $A_H = 0.4$\,aJ,
  van der Waals force = $-A_H\times R/(6(z+\Delta)^2)$, $\Delta=2$\,\AA,
  long-range electrostatic force = $U^2 \pi \epsilon _0 R/(z+ \Delta- \sigma)$.}
\label{gamvdW+est387}
\end{figure}}
Because we have approximated the charges of the dangling bonds to concentrate on a
point, the simulated adatom image shows two spherical subpeaks - in contrast to the
crescent shaped experimental subpeaks. We expect that if accurate charge distributions
were taken
into account, crescent shaped simulated images would result. The height of the simulated
adatoms, the distance of the subpeaks and the dip between the subpeaks is in excellent
agreement with the experimental images.

Since our first observation of subatomic features by AFM,
we have observed similar images of double peaked adatom structures in various other
experiments. In our first experiments, the orientation of the double
peaks was roughly perpendicular to the fast scanning direction. In Figure \ref{2peaks}
we present examples of AFM data where the orientation of the crescents is
clearly unrelated to the fast scanning
direction.
\parbox{17cm}{
\begin{figure}[htbp]
  \centering
  \centerline{\epsfxsize=8.4cm{\epsfbox{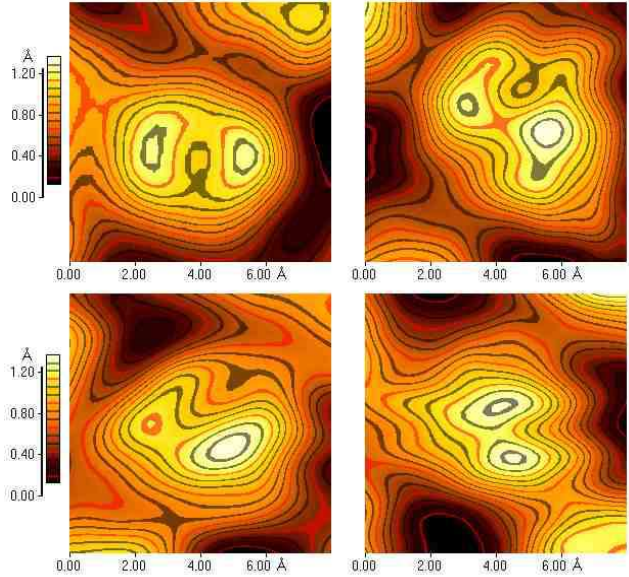}}}
  \caption{Experimental FM-AFM images of adatoms exposing
  two peaks per atom. In all the images,
  the fast scanning direction is horizontal ($x$). The orientation of the double
  peaks is clearly unrelated to the fast scanning direction.}\label{2peaks}
\end{figure}}

Because of the dramatic dependence of the images on the atomic tip state,
the preparation of AFM tips with a specified front atom symmetry and chemical identity
is instrumental for performing well defined experiments.
Commercial silicon cantilever tips are oriented in a (001)--direction.
In silicon, the natural cleavage planes are (111) planes. If a macroscopic tip
crystallite is limited by natural cleavage planes, a (001) oriented tip will in general
not end in a single atom tip, but in a rooftop type of symmetry. A tip which points into
a (111)--direction, however, will end in a single atom if we assume bulk termination.
Moreover, the front atom of this tip is expected to be particularly stable, because it is bonded
to the rest of the tip by three bonds and exposes a single dangling bond towards the sample.
Because of this prospects, we have attempted to cleave single crystalline silicon such that
we obtained crystallites which were limited by (111) planes supposedly ending in a single
atom tip \cite{Giessibl2001APA}. We have cleaved the silicon in air, so that an oxide layer of a typical
thickness of 20\,\AA{} will develop at the surface. As expected, the tips did not yield good
STM images after bringing them into vacuum. However, after exposing them to electron bombardment
(the same procedure is used for preparing the silicon 7$\times$7 surface), the tips work excellently
in STM mode and also in AFM mode. We speculate, that the (111) oriented sidewalls of the
tip will reconstruct just like a flat silicon surface. Even then, the front atom will
be bonded to three next neighbors and expose a single dangling bond.
Figure \ref{AFMi111} shows an AFM image obtained with such a (111) oriented tip.
\parbox{17cm}{
\begin{figure}[htbp]
  \centering
  \centerline{\epsfxsize=10cm{\epsfbox{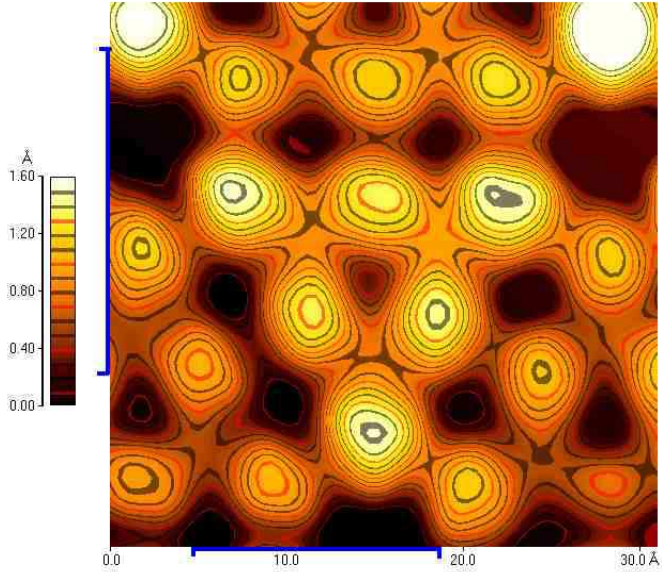}}}
  \caption{Topographic image of Si(111)-(7$\times$7) observed by FM-AFM, imaged with a
   single crystal silicon tip roughly oriented in a (111) direction.
   Imaging parameters: $k=1800$\,N/m,
   $A=2.5$\,\AA, $f_0=20531$\,Hz, $\Delta f=+85$\,Hz, thus $\gamma=+30$\,fNm$^{1/2}$,
   rms frequency error $\delta \Delta f=0.09$\,Hz, scanning speed = 20\,nm/s.
   The blue bars at the left and bottom border of the image indicate the $x,y$-range
   that is simulated in Fig. 16.}
   \label{AFMi111}
\end{figure}}
It is also noted, that this image was recorded at {\it positive}
frequency shift. Hence, the forces between front atom and sample
have been repulsive! An advantage of this mode is that the
feedback can be set much faster, because the risk of feedback
oscillations with a catastrophic tip crash is avoided in this
mode. Due to the faster feedback setting and the smaller scanning
speed, the relative rms frequency error (=$\delta \Delta f/\Delta
f$) is significantly smaller in Fig. \ref{AFMi111} (0.1 \%) than
in Fig. \ref{rtlltr} (5.6 \%). Figure \ref{111tipsim} shows the
simulated image. Because the forces are repulsive, the dominant
forces are modelled with the repulsive part of the SW potential.

\parbox{17cm}{
\begin{figure}[htbp]
  \centering
  \centerline{\epsfxsize=10cm{\epsfbox{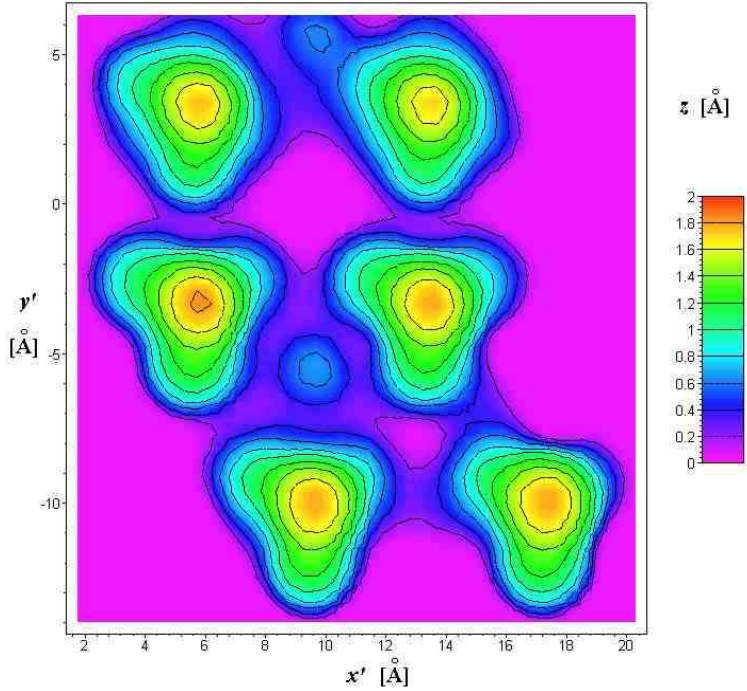}}}
  \caption{Simulated image of a section of the unit cell showing for $\gamma=+30$\,fNm$^{1/2}$. All four types of adatoms as well
  as the rest atoms are visible, imaged with a silicon crystallite oriented in a (111) direction with a front atom with a
  single dangling bond. The Stillinger-Weber potential has been used for the calculation of the image. The positions of the surface
  atoms are adapted from Tong {\it et al.} \protect\cite{Tong1988}.}
  \label{111tipsim}
\end{figure}}

\section{Discussion}

The experimental data which is presented here confirms the
prediction of Eq. \ref{z_noise}: the resolution of the AFM
is improved by working at small oscillation amplitudes. The resolution obtained in the AFM
images is high enough such that the angular dependence of the tip-sample forces is revealed.
Chemical and electrostatic force models have been employed to explain and analyze the experimental
images.
The choice of our model forces leads to a qualitative agreement between experiment and
simulations.
The double peak images are well explained by a SW potential and additional electrostatic
effects which polarize the orbitals. However, using only the SW potential for the short-range
contribution results in a simulated height of the adatoms which is only 10\,\% of the experimental value.
Adding the simple electrostatic short-range interaction, the height of the simulated adatoms agrees well
with experiment, while the shape of the adatom images differs from experiment.
The experimental images show crescents, while the
theoretical images are spherical.
It is apparent that the simulation of the images captures the basic principle, but needs to be refined
for obtaining a better agreement with experiment.

The simulations covering the repulsive regime with a SW potential leads to
a qualitative agreement between experiment and simulations. The simulation shows that the
rest atoms, located between a corner adatom and two center adatoms,
should be visible as a small protrusion. The experiment shows a saddlepoint
at the expected rest atom positions and a
hole between three center adatoms. However, the shape of the four types of adatoms is
similar in the simulation and very different in the experiment which shows that taking only next
neighbor interaction and bond angles into account is insufficient for a proper model of
the forces between silicon atoms. Also, the heights of the
adatoms is much different than measured by LEED or predicted by calculations \cite{Tong1988}.
According to LEED data, the corner adatoms are only about 0.04\,\AA{} higher than the center adatoms.
In our AFM data, the corner adatoms appear to be roughly 0.2 \,\AA{} higher than the center adatoms.
Because this height difference is measured both with repulsive and attractive short-range forces, it
cannot be caused by elastic deformations of tip and sample.

Our simulations have shown, that the observation of atomic orbitals is only possible if the
tip-sample distance is of the order of the interatomic distance in the bulk material.
The forces that act for such small distances are large, and deformations and increased
dissipation are expected to occur. These issues and the observability of atomic orbitals by STM
are discussed in the following subsections.

\subsection{Tip stability issues}

An important issue arises in the tip and sample stability. So far, we have assumed
that both tip and sample atoms are not strained when the tip comes close to the
sample. In STM, this might be the case when the tunneling impedance is large, and
even in dynamic AFM it is conceivable that $F_{ts}$ is small enough such that
the elastic deformation of tip and sample is not noticeable.

However, as is evident from Fig. \ref{sto}, the observation of deviations of the
spherical symmetry of atoms is only possible in the near
field, i.e. for distances which are of the order of the bulk interatomic distance.
Thus, for observing these features by AFM, tip and sample have to come very close,
and the atoms close to the interaction region will suffer from noticeable
strain. Figure \ref{100tip} shows a ball and
stick model of a silicon crystallite which is limited by (111) planes and points into
a (001) direction. This crystallite is used as a model for our tip.
\parbox{17cm}{
\begin{figure}[htbp]
  \centering
  \centerline{\epsfxsize=6cm{\epsfbox{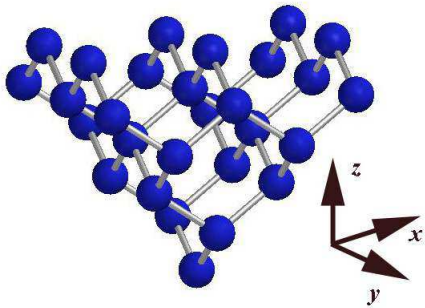}}}
  \caption{Model for a tip crystallite cleaved from bulk silicon with (111)-oriented
  surfaces and pointing in a (001) direction.}\label{100tip}
\end{figure}}

The stiffness of the bonds of the front atom to the upper part of
the tip can be estimated using the Stillinger-Weber potential. For the crystallite
shown in Fig. \ref{100tip}, the stiffness in $z$-direction is 115\,N/m,
in $x$-direction it is 20.4\,N/m and in $y$-direction it is 224\,N/m.
The stiffness can also be evaluated from the phonon spectrum of silicon.
For an optical phonon at the $\Gamma$-point (i.e. the wave vector is zero),
neighboring atoms vibrate opposite to each other.
We can model this situation with a single silicon atoms which is suspended by four
springs with half the length of the nearest neighbor distance, each having an effective
spring constant in $z$-direction of $k_{z\,half\,bond}.$
The wave number of optical
phonons in silicon at the $\Gamma$-point is 520\,cm$^{-1}$ which corresponds to
$f_{opt.\,phonon}\approx 15.589$\,THz \cite{Landolt1982}. The mean atomic mass
(natural mixture of isotopes) of silicon is
28.086\,amu \cite{Ashcroft}, thus the spring constant
$k_{z\,half\,bond}=\frac{1}{4}\times$447.45\,N/m = 111.9\,N/m.
The stiffness of the two bonds connecting the front atom of the tip crystallite with the rest of the
tip is given by twice the stiffness of one full-length bond with $k=\frac{1}{2}k_{z\,half\,bond}$
yielding $k_z= 111.9$ \,N/m -- in excellent agreement with the value of
115\,N/m derived from the Stillinger-Weber potential.

The positions
of the two atoms which attach to the front atom have been kept fixed for the derivations.
In reality, all the tip atoms will respond to the force on the front atom. The effect can be
estimated by arranging the springs corresponding to a tip atom layer in series.
For the tip shown in Fig. \ref{100tip}, the $m$-th bilayer contains one layer with $m^2$
and one layer with $m(m+1)$ atoms. Each of these atoms has two bonds to the layer above.
The effective
stiffness of the tip crystallite in $z$-direction is then given by
\begin{equation}
\frac{1}{k_{eff}} = \frac{1}{k_{z}} \sum_{m=1}^{\infty}[\frac{1} {m^2}+\frac{1} {m(m+1)}]=
\frac{1}{k_{z}} (1+\frac{\pi^2}{6})\approx \frac{1}{0.4\cdot k_{z}}.
\end{equation}
For tip-sample forces of the order of 2\,nN, we expect that the
bonds between the front atom of the tip and its next neighbors
will be strained by $2\,{\rm nN}/k_z\approx 0.2$\,\AA. The total
strain of the tip is expected to be $2\,{\rm nN}/k_z/0.4\approx
0.5$\,\AA. It is expected that the sample will become strained to
a similar extent.

\subsection{Amplitude feedback stability}

When the oscillating cantilever of an FM-AFM comes close to the sample, increased damping
has been observed experimentally \cite{Guggisberg2000,Luethi1997,Bammerlin1997,Ueyama1998}
and described theoretically \cite{Giessibl1999b,Durig2000,Durig1999,Gauthier1999,Abdurixit2000}.

We expect that in addition to the damping channels which are present in classic FM-AFM, an
even more significant damping channel will open up for very small tip sample distances.
When the front atom of the tip moves to the sample, both
the front atom and the adjacent sample atom will be pulled from their equilibrium
position and relax when the cantilever swings back from the sample.
This process will happen adiabatically and dissipate little energy as long as the
closest distance is larger than the distance
where the maximal attractive force $F_{att\,max}$ occurs (inflection point).

However, when the distance
between front atom and sample gets closer than the
inflection point, an additional channel of significant
energy loss will occur
because the front atom of the tip and the sample atom will be pulled out of
the equilibrium position, and rather than moving back adiabatically, will
stick together until the bonding force is overcome by the withdrawing cantilever.
At this point, the front atom and the sample atom will be pulled out by a
distance $F_{att\,max}/k_z$. When the cantilever swings back,
the front and sample atom will oscillate at the phonon frequency and dissipate
the stored energy
\begin{equation}
\Delta E = \frac{F_{att\,max}^2} {2\,k_z}.
\end{equation}
The maximal attractive force $F_{att\,max}$ is about
4\,nN, according to the SW potential.
Thus, the energy loss per cycle due to this process is 0.08\,aJ per atom. Because
both the front atom at the tip and the sample atom will be excited by this process, the total
energy loss is thus of the order of 1\,eV per cycle. This energy has to be provided by
the oscillator circuit which drives the cantilever. The intrinsic energy loss of the
cantilever is $\pi k A^2/Q$,
where $k$ is the spring constant of the cantilever, $A$ is its oscillation amplitude and
$Q$ is its quality factor. For the qPlus sensor, $Q\approx 4000$ and $k=1800$\,N/m. For
an amplitude of the order of 1\,nm, the intrinsic loss
per cycle is of the order of 10\,eV and the extra loss due to tip-sample dissipation is
small. With conventional silicon cantilevers, typical $Q$ factors are at least one order of
magnitude larger and the intrinsic dissipation is much smaller. Amplitude control can
become difficult with conventional cantilevers when attempting to get very
close to the sample as required for observing
the orbital structure of atoms.

\subsection{Observability of individual orbitals by STM}

To our knowledge, the observation of multiple maxima in the image of a single atom in
STM mode has not been reported before.
We can identify three possible explanations why the internal orbital electron structure
of atoms is observable by AFM, but apparently not by STM:

\begin{enumerate}
\item As shown in Fig. \ref{sto}, the tip sample distance needs to be very small before the
charge density of valence electrons displays a noticeable deviation from the rotational
symmetry about the $z-$ axis. Typical tip-sample distances (from the center of the tip atom
to the center of the closest sample atom) are $5-8$\,\AA{} in STM. When attempting to image
in STM mode at a tip sample distance of $2-3$\,\AA{}, the tip-sample force is very large and
the tip atom or sample atom may not withstand the shear forces acting during scanning. In dynamic
AFM, this small distance occurs only intermittently and the lateral movement can occur during
the oscillation phase where the tip is far from the sample;
\item While it is known that the force between two atoms is in general attractive at large
distances and becomes repulsive at short distances, the tunneling current is usually assumed to
be a monotonic function of the distance. Thus, it is expected that the orbital structure
of the valence electrons has a larger effect on the force than on the tunneling current;
\item In the STM mode, typical tunneling voltages used for imaging silicon are of the order of 2\,V and
the energy of the tip and sample states which contribute to the tunneling current
is spread over a range of 2\,eV. Just like in atomic wave functions,
where the angular dependent $p,d$ and $f$ sub-shells add up to a spherically symmetric charge
distribution if all the states of a given main quantum number are filled, we expect that a
similar process can happen in tunneling experiments when a range of tip and sample energies
contributes to the total tunneling current.
So it is conceivable that even if some
states which contribute to tunneling are not symmetric with respect to the $z-$axis, the sum of these
states is symmetric around $z$.
\end{enumerate}

In summary, we have shown that atomic force microscopy offers new insights into the
atomic structure and symmetry of surface atoms and AFM tips. We are confident that
using small oscillation amplitudes, stiff cantilevers and oscillator setups which are
insensitive to the enhanced damping at very close tip-sample distances provides a
new experimental tool to study the nature
of the atomic bond and the electronic structure of atoms.

%%%%%%%%%%%%%%%%%%%%%%%%%%%%%%%%%%%%%%%%%%%%%%%%%%%%%%%%%%%%%%%%%%%%%%%%%%
%%%%%%%%%%%%%%%%%%%%%% Acknowledgements %%%%%%%%%%%%%%%%%%%%%%%%%%%%%%%%%%
%%%%%%%%%%%%%%%%%%%%%%%%%%%%%%%%%%%%%%%%%%%%%%%%%%%%%%%%%%%%%%%%%%%%%%%%%%
\vspace*{0.25cm} \baselineskip=10pt{\small \noindent The authors are
  indebted to C. Laschinger and T. Kopp for discussions about the tip-sample
interaction, and to M. Lantz and H. Hug for fruitful conversations
about feedback stability. This work is supported by the BMBF
(project no. 13N6918/1).}
%%%%%%%%%%%%%%%%%%%%%%%%%%%%%%%%%%%%%%%%%%%%%%%%%%%%%%%%%%%%%%%%%%%%%%%%%%
%%%%%%%%%%%%%%%%%%%%%%%%% Bibliography %%%%%%%%%%%%%%%%%%%%%%%%%%%%%%%%%%%
%%%%%%%%%%%%%%%%%%%%%%%%%%%%%%%%%%%%%%%%%%%%%%%%%%%%%%%%%%%%%%%%%%%%%%%%%%
%%%%%%%%%%%%%% please use the following format for %%%%%%%%%%%%%%%%%%%%%%%
%%%%%%%%%%%%%%%%%%%%%%%%% JOURNAL REFERENCE %%%%%%%%%%%%%%%%%%%%%%%%%%%%%%
%
% \bibitem{journal} Author1, Author2, and Author3, Journal
%   {\bf Volume} (Year) Page_number
%
%%%%%%%%%%%%%%%%%%%%%%%%% BOOKS %%%%%%%%%%%%%%%%%%%%%%%%%%%%%%%%%%%%%%%%%%
%
% \bibitem{book} Author1, Author2, and Author3, {\it Book_Title},
%   Publisher, Place Year
%
%%%%%%%%%%%%%%%%%%% PROCEEDINGS AND OTHER EDITED BOOKS %%%%%%%%%%%%%%%%%%%
%
% \bibitem{edited} Author1 and Author2, in {\it Book_Title} edited by
%   Author3 and Author4, Publisher, Place Year, p. Page_Number
%
%%%%%%%%%%%%%%%%%%%%%%%%% THESIS %%%%%%%%%%%%%%%%%%%%%%%%%%%%%%%%%%%%%%%%%
%
% \bibitem{thesis} Author, PhD thesis, University, Place Year
%
%%%%%%%%%%%%%%%%%%%%%%%%% PREPRINT %%%%%%%%%%%%%%%%%%%%%%%%%%%%%%%%%%%%%%%
%
% \bibitem{preprint} Author1 and Author2, Preprint_name_and_number,
%   University, Place Year, submitted to Journal
%%%%% OR
%   accepted for publication in Journal
%%%%% OR
%   unpublished
%
%%%%%%%%%%%%%%%%%%%%%%%%% ELECTRONIC PREPRINT %%%%%%%%%%%%%%%%%%%%%%%%%%%%
%
% \bibitem{eprint} Author1, Author2, and Author3, cond-mat/9876543,
%   submitted to Journal
%%%%% OR
%   accepted for publication in Journal
%
%%%%%%%%%%%%%%%%%%%%%%%%%%%%%%%%%%%%%%%%%%%%%%%%%%%%%%%%%%%%%%%%%%%%%%%%%%

\end{document}